\documentstyle[aps]{revtex}

\begin{document}
\title{Final state phases in $B\rightarrow D\pi $, $\bar{D}\pi $ decays and
CP-asymmetry}
\author{Fayyazuddin}
\address{National Center for Physics\\
Quaid-i-Azam University\\
Islamabad 45320\\
Pakistan}
\date{[hep-ph/0402189]}
\maketitle

\begin{abstract}
Final state phases $\delta _f$ and $\delta _f^{\prime }$ in $B\rightarrow
D\pi $, $\bar{D}\pi $ decays are shown to be equal i.e.$\delta =\delta
_f-\delta _f^{\prime }=0$. Thus CP-violating asymmetry ${\cal A}\left(
t\right) $ is independent of final state phases. The estimate for the phases 
$\delta _f$ and $\delta _f^{\prime }$ is also given.
\end{abstract}

Time dependent $B$-decays are a good source of our knowledge regarding $CP$%
-violation. However $CP$-violation involves final state phases. Thus it is
not possible to extract the weak phase $\gamma $, without some knowledge of
final state phases. $\Delta C=\pm 1$, $\Delta S=0$ $B$-decays are of special
interest, because for these decays, it is possible to show that final state
phase $\delta =0$. The purpose of the paper is to show that this is the
case. It is convienent to write the time dependent decay rates in the form
(For a review see for example refrences \cite{r01,r02,r03}) 
\begin{eqnarray}
&&\left[ \Gamma \left( B^0\left( t\right) \rightarrow f\right) -\Gamma
\left( \bar{B}^0\left( t\right) \rightarrow \bar{f}\right) \right] +\left[
\Gamma \left( B^0\left( t\right) \rightarrow \bar{f}\right) -\Gamma \left( 
\bar{B}^0\left( t\right) \rightarrow f\right) \right]  \nonumber \\
&=&e^{-\Gamma t}\left\{ \cos \Delta mt\left[ \left( \left| \left\langle
f\left| H\right| B^0\right\rangle \right| ^2-\left| \left\langle \bar{f}%
\left| H\right| \bar{B}^0\right\rangle \right| ^2\right) +\left( \left|
\left\langle \bar{f}\left| H\right| B^0\right\rangle \right| ^2-\left|
\left\langle f\left| H\right| \bar{B}^0\right\rangle \right| ^2\right)
\right] \right.  \nonumber \\
&&\left. -2\sin \Delta mt\left[ 
\mathop{\rm Im}
\left( e^{2i\phi _M}\left\langle f\left| H\right| B^0\right\rangle
^{*}\left\langle f\left| H\right| \bar{B}^0\right\rangle \right) +%
\mathop{\rm Im}
\left( e^{2i\phi _M}\left\langle \bar{f}\left| H\right| B^0\right\rangle
^{*}\left\langle \bar{f}\left| H\right| \bar{B}^0\right\rangle \right)
\right] \right\}  \nonumber \\
&&  \label{e1} \\
&&\left[ \Gamma \left( B^0\left( t\right) \rightarrow f\right) +\Gamma
\left( \bar{B}^0\left( t\right) \rightarrow \bar{f}\right) \right] -\left[
\Gamma \left( B^0\left( t\right) \rightarrow \bar{f}\right) +\Gamma \left( 
\bar{B}^0\left( t\right) \rightarrow f\right) \right]  \nonumber \\
&=&e^{-\Gamma t}\left\{ \cos \Delta mt\left[ \left( \left| \left\langle
f\left| H\right| B^0\right\rangle \right| ^2+\left| \left\langle \bar{f}%
\left| H\right| \bar{B}^0\right\rangle \right| ^2\right) -\left( \left|
\left\langle \bar{f}\left| H\right| B^0\right\rangle \right| ^2+\left|
\left\langle f\left| H\right| \bar{B}^0\right\rangle \right| ^2\right)
\right] \right.  \nonumber \\
&&\left. -2\sin \Delta mt\left[ 
\mathop{\rm Im}
\left( e^{2i\phi _M}\left\langle f\left| H\right| B^0\right\rangle
^{*}\left\langle f\left| H\right| \bar{B}^0\right\rangle \right) -%
\mathop{\rm Im}
\left( e^{2i\phi _M}\left\langle \bar{f}\left| H\right| B^0\right\rangle
^{*}\left\langle \bar{f}\left| H\right| \bar{B}^0\right\rangle \right)
\right] \right\}  \nonumber \\
&&  \label{e2}
\end{eqnarray}
In refrence \cite{r04}, it was suggested that time dependent $B\rightarrow
D\pi $ decays can be used to find $\sin \left( 2\beta +\gamma \right) $. The
detailed analysis has been done in references \cite{r05,r06,r07}.

For $B\rightarrow D\pi $ decays, the decay amplitudes can be written as 
\begin{eqnarray}
A_{+-} &=&\left\langle \bar{f}\left| H\right| \bar{B}^0\right\rangle
=\left\langle D^{+}\pi ^{-}\left| H\right| \bar{B}^0\right\rangle =\bar{A}_{%
\bar{f}}  \nonumber \\
A_{-+} &=&\left\langle f\left| H\right| B^0\right\rangle =\left\langle
D^{-}\pi ^{+}\left| H\right| B^0\right\rangle =A_f\text{, }A_f=\bar{A}_{\bar{%
f}}  \nonumber \\
A_{-+}^{\prime } &=&\left\langle f\left| H\right| \bar{B}^0\right\rangle
=\left\langle D^{-}\pi ^{+}\left| H\right| \bar{B}^0\right\rangle
=e^{i\gamma }\bar{A}_f^{\prime }  \nonumber \\
A_{+-}^{\prime } &=&\left\langle \bar{f}\left| H\right| B^0\right\rangle
=\left\langle D^{+}\pi ^{-}\left| H\right| B^0\right\rangle =e^{-i\gamma }%
\bar{A}_f^{\prime }  \label{e3}
\end{eqnarray}
Note that the effective Lagrangians for decays $\bar{B}^0\rightarrow
D^{+}\pi ^{-}$ and $\bar{B}^0\rightarrow D^{-}\pi ^{+}$ are given by 
\begin{mathletters}
\begin{eqnarray}
&&V_{cb}V_{ud}^{*}\left[ \bar{d}\gamma ^\mu \left( 1+\gamma _5\right)
u\right] \left[ \bar{c}\gamma _\mu \left( 1+\gamma _5\right) b\right]
\label{e4a} \\
&&V_{ub}V_{cd}^{*}\left[ \bar{d}\gamma ^\mu \left( 1+\gamma _5\right)
c\right] \left[ \bar{u}\gamma _\mu \left( 1+\gamma _5\right) b\right]
\label{e4b}
\end{eqnarray}
respectively. In the Wolfenstein parametrization of $CKM$ matrix 
\end{mathletters}
\begin{equation}
\frac{V_{cb}V_{ud}^{*}}{V_{ub}V_{cd}^{*}}=\lambda ^2\sqrt{\rho ^2+\eta ^2}%
e^{i\gamma }  \label{e5}
\end{equation}
Thus for $B\rightarrow D\pi $ decays, we get from Eqs. (\ref{e1}), (\ref{e2}%
) and (\ref{e3}) 
\begin{eqnarray}
{\cal A}\left( t\right) &\equiv &\frac{\left[ \Gamma \left( B^0\left(
t\right) \rightarrow f\right) -\Gamma \left( \bar{B}^0\left( t\right)
\rightarrow \bar{f}\right) \right] }{\left[ \Gamma \left( B^0\left( t\right)
\rightarrow f\right) +\Gamma \left( \bar{B}^0\left( t\right) \rightarrow
f\right) \right] }  \nonumber \\
&=&-\sin \Delta mt\sin \left( 2\beta +\gamma \right) \frac{A_f^{*}\bar{A}%
_f^{\prime }+A_f\bar{A}_f^{\prime *}}{\left| A_f\right| ^2+\left| \bar{A}%
_f^{\prime }\right| ^2}  \nonumber \\
&=&-\frac{2r}{1+r^2}\sin \Delta mt\cos \left( \delta _f-\delta _f^{\prime
}\right) \sin \left( 2\beta +\gamma \right)  \label{e6} \\
{\cal F}\left( t\right) &\equiv &\frac{\left[ \Gamma \left( B^0\left(
t\right) \rightarrow f\right) +\Gamma \left( \bar{B}^0\left( t\right)
\rightarrow \bar{f}\right) \right] -\left[ \Gamma \left( B^0\left( t\right)
\rightarrow \bar{f}\right) +\Gamma \left( \bar{B}^0\left( t\right)
\rightarrow f\right) \right] }{\left[ \Gamma \left( B^0\left( t\right)
\rightarrow f\right) +\Gamma \left( \bar{B}^0\left( t\right) \rightarrow 
\bar{f}\right) \right] +\left[ \Gamma \left( B^0\left( t\right) \rightarrow 
\bar{f}\right) +\Gamma \left( \bar{B}^0\left( t\right) \rightarrow f\right)
\right] }  \nonumber \\
&=&\frac{\left| A_f\right| ^2-\left| \bar{A}_f^{\prime }\right| ^2}{\left|
A_f\right| ^2+\left| \bar{A}_f^{\prime }\right| ^2}\cos \Delta mt+i\frac{%
A_f^{*}\bar{A}_f^{\prime }-A_f\bar{A}_f^{\prime *}}{\left| A_f\right|
^2+\left| \bar{A}_f^{\prime }\right| ^2}\sin \Delta mt\cos \left( 2\beta
+\gamma \right)  \nonumber \\
&=&\frac{1-r^2}{1+r^2}\cos \Delta mt-\frac{2r}{1+r^2}\sin \Delta mt\cos
\left( 2\beta +\gamma \right) \sin \left( \delta _f-\delta _f^{\prime
}\right)  \label{e7}
\end{eqnarray}
where $r=\left| \bar{A}_f^{\prime }\right| ^2/\left| A_f\right| ^2$ and $f$
and $\bar{f}$ stand for $D^{-}\pi ^{+}$ and $D^{+}\pi ^{-}$ respectively. It
is clear from Eqs. (\ref{e6}) and (\ref{e7}), that extraction of angle $%
2\beta +\gamma $ depends on the strong phases of amplitude $A_f$ and $\bar{A}%
_f^{\prime }$ and parameter $r$.

The rest of the paper is concerned with the strong phases. First we consider
the decays 
\begin{eqnarray*}
\bar{B}^0\left( t\right)  &\rightarrow &D^{+}\pi ^{-} \\
&\rightarrow &D^0\pi ^0 \\
B^{-}\left( t\right)  &\rightarrow &D^0\pi ^{-}
\end{eqnarray*}
The effective Lagrangian (\ref{e4a}) for these decays has $\Delta I=1$. As
is well known for these decays, isospin analysis gives 
\begin{mathletters}
\label{e8}
\begin{eqnarray}
A_{+-} &=&\frac 13\left[ A_{3/2}+2A_{1/2}\right] =\frac 13\left[
f_{3/2}e^{i\delta _{3/2}}+2f_{1/2}e^{i\delta _{1/2}}\right] :T+A_2
\label{e8a} \\
A_{00} &=&-\frac{\sqrt{2}}3\left[ A_{3/2}-A_{1/2}\right] =-\frac{\sqrt{2}}3%
\left[ f_{3/2}e^{i\delta _{3/2}}-f_{1/2}e^{i\delta _{1/2}}\right] :-\frac 1{%
\sqrt{2}}\left( C-A_2\right)   \label{e8b} \\
A_{0-} &=&A_{3/2}=f_{3/2}e^{i\delta _{3/2}}:T+C,  \label{e8c} \\
A_{+-}-\sqrt{2}A_{00} &=&A_{0-}  \label{e9}
\end{eqnarray}
where $T,C,A_2$ denote contributions from the tree, the color suppressed and 
$W$-exchange diagrams respectively.

On the other hand for the effective Lagrangian (\ref{e4b}), we have both $%
\Delta I=1$ and $\Delta I=0$ parts. Thus for the decays 
\end{mathletters}
\begin{eqnarray*}
\bar{B}^0\left( t\right)  &\rightarrow &D^{-}\pi ^{+} \\
&\rightarrow &\bar{D}^0\pi ^0 \\
B^{-}\left( t\right)  &\rightarrow &D^{-}\pi ^0 \\
&\rightarrow &\bar{D}^0\pi ^{-}
\end{eqnarray*}
the isospin analysis gives 
\begin{mathletters}
\label{e8}
\begin{eqnarray}
A_{-+}^{\prime } &=&-\frac{\sqrt{2}}3A_{3/2}^{\prime }+\frac{\sqrt{2}}3%
C_{1/2}^{\prime }-\sqrt{\frac 23}D_{1/2}^{\prime }:T^{\prime }+A_2^{\prime }
\label{e10a} \\
A_{00}^{\prime } &=&\frac 23A_{3/2}^{\prime }+\frac 13C_{1/2}^{\prime }-%
\frac 1{\sqrt{3}}D_{1/2}^{\prime }:-\frac 1{\sqrt{2}}\left( C^{\prime
}-A_2^{\prime }\right)   \label{e10b} \\
A_{-0}^{\prime } &=&-\frac 23A_{3/2}^{\prime }-\frac 13C_{1/2}^{\prime }-%
\frac 1{\sqrt{3}}D_{1/2}^{\prime }:\frac 1{\sqrt{2}}\left( T^{\prime
}-A_1^{\prime }\right)   \label{e10c} \\
A_{0-}^{\prime } &=&-\frac{\sqrt{2}}3A_{3/2}^{\prime }+\frac{\sqrt{2}}3%
C_{1/2}^{\prime }+\sqrt{\frac 23}D_{1/2}^{\prime }:C^{\prime }+A_1^{\prime }
\label{e10d}
\end{eqnarray}
where $D_{1/2}^{\prime }$ is the contribution from $\Delta I=0$ part of the
effective Lagrangian. Here it is convenient to write 
\end{mathletters}
\begin{eqnarray}
C_{1/2}^{\prime }-\sqrt{3}D_{1/2}^{\prime } &=&A_{1/2}^{\prime }  \nonumber
\\
C_{1/2}^{\prime }+\sqrt{3}D_{1/2}^{\prime } &=&B_{1/2}^{\prime }  \label{e12}
\end{eqnarray}

Thus we can write 
\begin{mathletters}
\label{e13}
\begin{eqnarray}
A_{-+}^{\prime } &=&-\frac{\sqrt{2}}3\left[ A_{3/2}^{\prime
}-A_{1/2}^{\prime }\right] =-\frac{\sqrt{2}}3\left[ f_{3/2}^{\prime
}e^{i\delta _{3/2}^{\prime }}-f_{1/2}^{\prime }e^{i\delta _{1/2}^{\prime
}}\right]  \label{e13a} \\
A_{00}^{\prime } &=&\frac 13\left[ 2A_{3/2}^{\prime }+A_{1/2}^{\prime
}\right] =\frac 13\left[ 2f_{3/2}^{\prime }e^{i\delta _{3/2}^{\prime
}}+f_{1/2}^{\prime }e^{i\delta _{1/2}^{\prime }}\right]  \label{e13b} \\
A_{-0}^{\prime } &=&-\frac 13\left[ 2A_{3/2}^{\prime }+B_{1/2}^{\prime
}\right] =-\frac 13\left[ 2f_{3/2}^{\prime }e^{i\delta _{3/2}^{\prime
}}+g_{1/2}^{\prime }e^{i\delta _{1/2}^{\prime \prime }}\right]  \label{e13c}
\\
A_{0-}^{\prime } &=&-\frac{\sqrt{2}}3\left[ A_{3/2}^{\prime
}-B_{1/2}^{\prime }\right] =-\frac{\sqrt{2}}3\left[ f_{3/2}^{\prime
}e^{i\delta _{3/2}^{\prime }}-g_{1/2}^{\prime }e^{i\delta _{1/2}^{\prime
\prime }w}\right]  \label{e13d} \\
A_{-+}^{\prime }-\sqrt{2}A_{00}^{\prime } &=&\sqrt{2}A_{-0}^{\prime
}+A_{0-}^{\prime }  \label{e13e}
\end{eqnarray}
In order to calculate the final state phases $\delta $'s, the following
physical picture is useful. In the weak decays of $B$-mesons, the $b$-quark
is converted into $b\rightarrow cq\bar{q}$, $b\rightarrow uq\bar{q};$ since
for the \cite{r02,08,10,11} tree graph the configration is such that $q$ and 
$\bar{q}$ essentially go togather into a color singlet state with the third
quark recoiling, there is a signifcant probability that the system will
hadronize as a two body final state. Thus the strong phase shifts are
expected to be small at least for tree amplitude. They are generated after
hydronization by rescattering.

As noted in Ref. \cite{12}, in the simple factorization ansatz (large $N_c$
limit), $C$ and $A_2$ vanishes so that the decay $\bar{B}^0\rightarrow
D^0\pi ^0$ entirely arises from rescattering: 
\end{mathletters}
\[
\bar{B}^0\rightarrow D^{+}\pi ^{-}\rightarrow D^0\pi ^0 
\]
Thus, the rescattering involves charge exchange; hence it is determined by
an isospin one exchange trajectory. The isospin decomposition of the
scattering amplitude $\pi _iD\rightarrow \pi _jD$ can be expressed in terms
of two amplitudes 
\begin{equation}
M_{ji}=M^{\left( +\right) }\delta _{ji}+\frac 12M^{\left( -\right) }\left[
\tau _j,\tau _i\right]
\end{equation}
The $I=3/2$ and $I=1/2$ scattering amplitudes are related to $M^{\left(
+\right) }$ and $M^{\left( -\right) }$ as follows 
\begin{eqnarray}
M_{3/2} &=&M^{\left( +\right) }-M^{\left( -\right) }  \nonumber \\
M_{1/2} &=&M^{\left( +\right) }+2M^{\left( -\right) }  \nonumber
\end{eqnarray}
Since an isospin one exchange contributes to $M^{\left( -\right) }$, we get 
\cite{12} 
\begin{eqnarray}
\frac{M_{3/2}}{M_{1/2}}=-\frac 12
\end{eqnarray}
In order to consider the rescattering corrections to the decays $\bar{B}%
^0\rightarrow D^0\pi ^0$ and $\bar{B}^0\rightarrow \bar{D}^0\pi ^0$, we note
that the unitarity gives the imaginary part of the decay amplitude $A_f$ as
follows: 
\begin{equation}
\mathop{\rm Im}
A_f=\sum_nM_{nf}^{*}A_n  \label{ns14}
\end{equation}
where $M_{nf}$ is the scattering amplitude for $f\rightarrow n$ and $A_n$ is
the decay amplitude $\bar{B}^0\rightarrow n$. In dispersion relation two
particle unitarity gives dominant contribution. The dominant contribution is
expected to be from the intermediate state $n=f^{\prime }$ $(f^{\prime }=\pi
^{-}D^{+}$ or $\pi ^{+}D^{-})$. Other possible two particle states which may
contribute are $\pi ^{-}D^{*+}(\pi ^{+}D^{*-})$ or $\rho ^{-}D^{+}\left(
\rho ^{+}D^{-}\right) $. But it is important to note that $\bar{B}%
^0\rightarrow \pi ^{-}D^{+}(\pi ^{+}D^{-})$ decays are s-wave decays whereas 
$\pi ^{-}D^{*+}(\pi ^{+}D^{*-})$ are p-wave decays. Hence the scatttering
amplitude connecting p-wave intermediate states with s-wave final states in
Eq. (\ref{ns14}) will not contribute. Similar remarks hold for $\rho
^{-}D^{+}\left( \rho ^{+}D^{-}\right) $ states.

Now the scattering amplitude $M_{ff^{\prime }}$ is given by $I=1$ ($\rho $
trajectory) exchange in the $t$-channel. Then using unsubtracted dispersion
for the decay amplitude $A_f$, the rescattering correction to $A_f$ can be
written in the form $\epsilon e^{i\theta }A_{f^{\prime }}$ \cite{13,14,15}.

Hence taking into account the rescattering correction, Eqs. (\ref{e8b}) and (%
\ref{e13b}) are modifird to 
\begin{eqnarray}
A_{00} &=&-\frac 1{\sqrt{2}}\left( C-A_2\right) -\sqrt{2}\epsilon e^{i\theta
}T\text{,}  \label{ns15} \\
A_{00}^{\prime } &=&-\frac 1{\sqrt{2}}\left( C^{\prime }-A_2^{\prime
}\right) -\sqrt{2}\epsilon e^{i\theta }T^{\prime }  \label{ns16}
\end{eqnarray}
Note an important fact that $I=1$ exchange in the $t$-channel gives the same
contribution to rescattering amplitudes both for $\pi D$ and $\pi \bar{D}$
channels. This is a consequence of $C$-invariance of scattering amplitude: 
\begin{eqnarray}
M_{ff^{\prime }} &=&\left\langle \pi ^0D^0\left| M\right| \pi
^{-}D^{+}\right\rangle =\left\langle \pi ^0D^0\left| C^{-1}CMC^{-1}C\right|
\pi ^{-}D^{+}\right\rangle  \nonumber \\
&=&\left\langle \pi ^0\bar{D}^0\left| M\right| \pi ^{+}D^{-}\right\rangle
\label{ns17}
\end{eqnarray}

First we note that the strong phases are generated after hadronization i.e.
they are negligibly small in the absence of rescattering. Hence in the
absence of rescattering 
\begin{equation}
A_{00}=-\frac{\sqrt{2}}3\left[ f_{3/2}-f_{1/2}\right] =-\frac 1{\sqrt{2}}%
\left( C-A_2\right) =-\frac 1{\sqrt{2}}\left( 2b\right) T  \label{ns18}
\end{equation}
where 
\begin{equation}
2b=\frac{C-A_2}T  \label{ns19}
\end{equation}
Now in high $N_c$ limit $\left( C-A_2\right) \rightarrow 0$, $%
f_{3/2}=f_{1/2} $. But for finite but small $b$, we get from Eq. (\ref{ns18}%
) 
\begin{equation}
\frac{f_{1/2}}{f_{3/2}}=\frac{\left( 1-b\right) }{1+2b}  \label{ns20}
\end{equation}
Similarly, we get 
\begin{equation}
-\frac{f_{1/2}^{\prime }}{2f_{3/2}^{\prime }}=\frac{\left( 1-b\right) }{1+2b}
\label{ns21}
\end{equation}
Secondly, after rescattering correction, $A_{00}$ is given by (cf. Eq. (\ref
{ns15})) 
\begin{equation}
A_{00}=-\frac{\sqrt{2}}3\left[ \left( f_{3/2}-f_{1/2}\right) -\epsilon
e^{i\theta }\left( f_{3/2}+2f_{1/2}\right) \right]  \label{ns22}
\end{equation}
We note that in the presence of rescattering $A_{00}$ is modified and can be
written in the form 
\begin{equation}
A_{00}=-\frac{\sqrt{2}}3\left[ \tilde{f}_{3/2}e^{i\delta _{3/2}}-\tilde{f}%
_{1/2}e^{i\delta _{1/2}}\right]  \label{ns23}
\end{equation}
Comparison with Eq. (\ref{ns22}) gives 
\begin{eqnarray}
\tilde{f}_{3/2} &=&f_{3/2}\left( 1+\epsilon \cos \theta \right) ,\,\,\,%
\tilde{f}_{1/2}=f_{1/2}\left( 1-2\epsilon \cos \theta \right)  \label{ns24}
\\
\tilde{f}_{3/2}\delta _{3/2}-\tilde{f}_{1/2}\delta _{1/2} &=&\epsilon \sin
\theta \left( f_{3/2}+2f_{1/2}\right)  \label{ns25}
\end{eqnarray}
Neglecting the terms of order $\epsilon \delta $, we get from Eq. (\ref{ns25}%
) 
\begin{equation}
f_{3/2}\delta _{3/2}-f_{1/2}\delta _{1/2}=\epsilon \sin \theta \left(
f_{3/2}+2f_{1/2}\right)  \label{ns26}
\end{equation}
From Eq. (\ref{ns26}), using Eq. (\ref{ns20}), we get 
\begin{equation}
\left( \delta _{3/2}-\delta _{1/2}\right) +b\left( 2\delta _{3/2}+\delta
_{1/2}\right) =3\epsilon \sin \theta  \label{ns27}
\end{equation}
Since right hand side of Eq. (\ref{ns27}) is independent of $b$, we obtain 
\begin{equation}
\delta _{1/2}=-2\delta _{3/2},\,\,\delta _{3/2}=\epsilon \sin \theta
\label{ns28}
\end{equation}
Similarly we get 
\begin{equation}
2f_{3/2}^{\prime }\delta _{3/2}^{\prime }+f_{1/2}^{\prime }\delta
_{1/2}^{\prime }=2\epsilon \sin \theta \left( f_{3/2}^{\prime
}-f_{1/2}^{\prime }\right)  \label{ns29}
\end{equation}
Then using Eq. (\ref{ns21}), we get 
\begin{equation}
\delta _{1/2}^{\prime }=-2\delta _{3/2}^{\prime },\,\,\delta _{3/2}^{\prime
}=\epsilon \sin \theta  \label{ns30}
\end{equation}
Eqs. (\ref{ns28}) and (\ref{ns30}) are our main results. From these results,
it follows that $\delta _f^{\prime }=\delta _f$ [see below].

Hence, we get from Eq.(6) and (7), 
\begin{eqnarray}
{\cal A}\left( t\right) &=&-\frac{2r}{1+r^2}\sin \Delta mt\sin (2\beta
+\gamma )  \label{ns31} \\
{\cal F}\left( t\right) &=&\frac{1-r^2}{1+r^2}\cos \Delta mt  \label{ns32}
\end{eqnarray}
It is clear from Eqs. (\ref{ns31}) and (\ref{ns32}) that $CP$-asymmetry is
independent of final state phase $\delta =\delta _f-\delta _f^{\prime }$.
Thus both the phase $2\beta +\gamma $ and the ratio $r$ can be determined
from the experimental values of ${\cal A}\left( t\right) $ and ${\cal F}%
\left( t\right) $. This result is also relevant for the recent measurement
of time-dependent $CP$ asymmetries for the $B^0\rightarrow D^{*\stackrel{-}{+%
}}\pi ^{\stackrel{+}{-}}$decays by BaBar\ collaboration \cite{18}. With $%
\delta =0$, their measurement of $CP$-asymmetries contain only one parameter 
$r$, instead of two for determining $\sin (2\beta +\gamma \dot{)}$.

Now from Eq. (8), using Eqs. (\ref{ns20}) and (\ref{ns28}), we obtain 
\begin{eqnarray}
A_f &\equiv &A_{+-}=f_{3/2}\frac 1{1+2b}e^{-i\delta _f}\equiv \left|
A_f\right| e^{-i\delta _f}  \nonumber \\
A_{00} &=&-\sqrt{2}\times f_{3/2}\frac 1{1+2b}(a+i\delta _{3/2})  \nonumber
\\
A_{0-} &=&f_{3/2}e^{i\delta _{3/2}}  \label{ns33}
\end{eqnarray}
where 
\begin{eqnarray}
\delta _f &=&\left( 1-2b\right) \delta _{3/2}  \label{ns34} \\
a &=&b+\epsilon \cos \theta  \label{ns35}
\end{eqnarray}
Similarly, from Eq. (11a), using Eqs. (\ref{ns21}) and (\ref{ns30}), we get 
\begin{eqnarray}
\bar{A}_f^{\prime } &\equiv &A_{-+}^{\prime }=-\frac{\sqrt{2}}{1+2b}%
f_{3/2}^{\prime }e^{-i\delta _f^{\prime }}  \nonumber \\
&\equiv &\left| \bar{A}_f^{\prime }\right| e^{-i\delta _f^{\prime }}
\label{ns36}
\end{eqnarray}
where 
\begin{equation}
\delta _f^{\prime }=\left( 1-2b\right) \delta _{3/2}^{\prime }=\delta _f
\label{ns37}
\end{equation}
Finally, we get from Eqs. (\ref{ns33}), 
\begin{equation}
R_{+-}\equiv \frac{\Gamma (\bar{B}^0\rightarrow D^{+}\pi ^{-})}{\Gamma
(B^{-}\rightarrow D^0\pi ^{-})}=\frac 1{(1+2b)^2}\approx 0.57\pm 0.09
\label{ns38}
\end{equation}
\begin{equation}
R_{00}\equiv \frac{\Gamma (\bar{B}^0\rightarrow D^0\pi ^0)}{\Gamma
(B^{-}\rightarrow D^0\pi ^{-})}=\frac 2{(1+2b)^2}(a^2+\delta
_{3/2}^2)\approx 0.06\pm 0.01  \label{ns39}
\end{equation}
where the numbers on the right hand side of Eqs. (\ref{ns38}) and (\ref{ns39}%
) are experimental values \cite{c17}.

From the above equations, we obtain 
\begin{eqnarray}
b &\approx &0.162  \label{ns40} \\
a^2+\delta _{3/2}^2 &=&0.053  \nonumber
\end{eqnarray}
Further we note that 
\begin{equation}
\delta _{3/2}=\epsilon \sin \theta \leq \epsilon  \label{ns41}
\end{equation}
Thus we cannot get the magnitude of phase shift unless we know $\theta $.
But it is expected to be small, since $\epsilon \sim 1/m_B$. In particular
for 
\begin{eqnarray}
\theta &=&90^0,\,\,\,\delta _{3/2}=\epsilon ,\,\,\,a=b,  \label{ns42} \\
\delta _{3/2} &\approx &0.162\approx 9^0  \label{ns43}
\end{eqnarray}
To conclude based on the assumption that final state phase shifts are small;
they are generated by rescattering, we have obtained $\delta _{3/2}^{\prime
}=\delta _{3/2}$, $\delta _{1/2}^{\prime }=\delta _{1/2}=-2\delta _{3/2}$, $%
\delta _f^{\prime }=\delta _f$. The equality of phase shifts for $\pi D$ and 
$\pi \bar{D}$ channels (which are $C$-conjugate of each other) is
essentially a consequence of $C$-invariance of scattering amplitude $M_{nf}$
for these channels.

Acknowledgement: This work was supported in parts by the Pakistan Council
for Science and Technology.

\end{document}